\documentclass[aps,prb,twocolumn, showpacs,notitlepage]{revtex4-1}

\usepackage{amsmath}
\usepackage{graphicx}
\usepackage{lmodern}
\usepackage{amsmath}
\usepackage{bm}
\usepackage{hyperref}\usepackage{url}
\usepackage{subfigure}%
\usepackage{dsfont}
\usepackage[normalem]{ulem}

\usepackage{amsbsy}
\usepackage{dcolumn}
\usepackage{amsthm}
\usepackage{bm}
\usepackage{esint}
\usepackage{multirow}
\usepackage{hyperref}
\usepackage{ulem}

\usepackage{cleveref}

\usepackage{mathrsfs}
\usepackage{amsfonts}
\usepackage{amsbsy}
\usepackage{dcolumn}
\usepackage{bm}
\usepackage{multirow}
\usepackage{color}

\usepackage{color}
\usepackage{amssymb}
\renewcommand{\vec}[1]{\boldsymbol{\mathbf{#1}}}
\newcommand{\beq} {\begin{equation}}
\newcommand{\eeq} {\end{equation}}
\newcommand{\bea} {\begin{eqnarray}}
\newcommand{\eea} {\end{eqnarray}}
\newcommand{\be} {\begin{equation}}
\newcommand{\ee} {\end{equation}}
\renewcommand{\(}{\left(}
\renewcommand{\)}{\right)}
\renewcommand{\[}{\left[}
\renewcommand{\]}{\right]}

\DeclareMathOperator{\sgn}{sgn}
\DeclareMathOperator{\Tr}{Tr}

\begin{document}

\title {Weak-Pairing Higher Order Topological Superconductors}
\author{Yuxuan Wang}
\author{Mao Lin}
\author{Taylor L. Hughes}
\affiliation{Department of Physics and Institute for Condensed Matter Theory, University of Illinois at Urbana-Champaign, 1110 West Green Street, Urbana, Illinois 61801-3080, USA }


\begin{abstract}
Conventional topological superconductors are fully gapped in the bulk but host gapless Majorana modes on their boundaries. We instead focus on a new class of superconductors, second-order topological superconductors, that have gapped, topological surfaces and gapless Majorana modes instead on lower-dimensional boundaries, i.e., corners of a two-dimensional system or hinges for a three-dimensional system. 
Here we propose two general scenarios in which second-order topological superconductivity can be realized spontaneously with weak-pairing instabilities. First, we show that $p_x+ip_y$-wave pairing in a (doped) Dirac semimetal in two dimensions with four mirror symmetric Dirac nodes realizes second-order topological superconductivity. Second, we show that $p+id$ pairing on an ordinary spin-degenerate Fermi sruface realizes second-order topological superconductivity as well. In the latter case we find that the topological invariants describing the system can be written using simple formulae involving only the low-energy properties of the Fermi surfaces and superconducting pairing. In both cases we show that these exotic superconducting states can be intrinsically realized in a metallic system with electronic interactions. For the latter case we also show it can be induced by proximity effect {in a heterostructure of cuprate and topological superconductors.}
\end{abstract}
\date{\today}

\maketitle

\section{Introduction}

 One of the characteristic properties of topological insulators (TIs) and superconductors (TSCs) is the presence of  stable, gapless modes hosted on their boundaries.  Such surface states are special because they cannot be realized in their intrinsic dimension having the same symmetries. A well-known example is the one-dimensional (1d) $p$-wave superconducting wire that is gapped in the bulk, but exhibits Majorana zero mode bound states (MBS) localized at its two ends~\cite{kitaev2001}. In higher dimensions, there are a wide variety of phases including 2d Chern insulators that host chiral edge states~\cite{haldane1988}, and 3d time-reversal invariant topological insulators that exhibit an odd-number of surface Dirac cones~\cite{hasankane,Qi2011}. The topological boundary modes  are  commonly used to diagnose the presence of the topological phase, e.g., by identifying the surface Dirac cone spectrum of topological insulators through angular resolved photoemission spectroscopy~\cite{hasankane,Qi2011}. They also generate much of the intrinsic interest in these systems for possible applications, e.g., using MBS as topological qubits~\cite{tqcreview}, or chiral modes as dissipationless transport channels. 
 
Recently, the notion of topological insulators has been extended to include higher-order topological insulators\cite{benalcazar2017,benalcazarPRB,schindler2017,song2017d,langbehn2017reflection,piet-new}; a new class of topological phases without  gapless surface states. A 2nd order topological insulator/superconductor (TI$_2$/TSC$_2$) is a $d$-dimensional system with \emph{gapped} $(d-1)$-dimensional boundaries that are themselves topologically non-trivial such that there are protected low-energy modes at the $(d-2)$-dimensional boundaries, e.g., corners in 2d and hinges in 3d. The first predicted TI$_2$ is the 2d quantized electric quadrupole insulator~\cite{benalcazar2017,benalcazarPRB,schindler2017,song2017d,langbehn2017reflection,piet-new} that has gapped edge states, but hosts  degenerate low-energy modes localized at the corners of a sample. This topological phase can be protected by a variety of symmetries, but the most commonly considered ones are either a pair of mirror symmetries $M_x, M_y,$ or $C_4$ symmetry. A simple model for this phase was proposed in Ref.~\onlinecite{benalcazar2017}, and was subsequently realized experimentally in three independent meta-material contexts~\cite{huber2018,bahl2018,thomale2018}.

 
 
In this article, we focus on higher-order topological superconductors~\cite{Brouwer2017,Khalaf2018,Felix2017,langbehn2017reflection,song2017d,wang-nandkishore-2017,Ryu2018,broiwer-2018,zhongwang-new}.  In analogy with 2d TI$_2$s, we provide mean-field Bogoliubov-de-Gennes (BdG) Hamiltonians that exhibit second-order topological superconducting phases and stable corner MBS. Corner Majorana state has been predicted in other superconductors with defects\cite{Teo2013}, such as impurity\cite{Slager2015} and distillations\cite{Benalcazar2014}. We explore two general scenarios in which one can \emph{spontaneously} realize TSC$_2$s. First,  we focus on mirror-symmetries and show that for a normal state corresponding to a two-dimensional Dirac semimetal with four mirror symmetric Dirac nodes, a $p_x+ip_y$ order parameter will generate second-order topology. Typically, a $p_x+ip_y$-wave superconducting order parameter gives rise to a Chern number and associated chiral Majorana edge modes. Here, however, something unusual happens due to the normal-state electronic structure, and the $p_x+ip_y$ order does not induce a nonzero Chern number, instead producing a TSC$_2$ with a $\mathbb{Z}_2$ topological invariant protected by mirror or particle-hole symmetries.  We consider the effects of shifting the position of the Dirac nodes, gapping them out, and doping them and find that the TSC$_2$ phase remains robust in a wide range of parameter space. Moving the Dirac nodes (in a mirror symmetric fashion) does not change the topology as long as they do not collapse and annihilate. Gapping out the Dirac points competes with the SC order parameter, but we show that the topology and corner MBS are robust as long as the Dirac mass is smaller than the superconducting order parameter. Additionally, with a finite chemical potential, the Dirac points in the normal state evolve into Fermi surfaces, and can pass through a Lifshitz transition to eventually shrink and vanish.  We show that the topology of the superconducting state remains robust throughout this process until the Fermi surfaces vanish.

The second context we consider is based on $C_4\mathcal{T}$ symmetry in a system realizing $p+id$ superconductivity. The $d$-wave order is odd under $C_4$ lattice rotation symmetry, and its $\pi/2$ relative phase to the $p$-wave order ensures a combined $C_4\mathcal{T}$ symmetry. The normal state in this case is a featureless, spin-degenerate Fermi surface. To understand the origin of the topological phase heuristically one can think of a two-stage process where the normal metal state forms a nodal $d$-wave superconductor with four nodal points, and then the nodal BdG quasiparticles are fully gapped by the generation of coexisting $p$-wave superconductivity. This process could arise in, e.g., $d$-wave cuprate superconductors when co-existing $p$-wave order is intrinsically or extrinsically/proximity induced. Alternatively, one could start from a normal state that is first fully gapped in the bulk by  $p_x+ip_y/p_x-ip_y$ pairing to form a time-reversal invariant TSC.~\cite{schnyder,raghu} As such, the system will have protected edge states and the addition of  $d$-wave order can gap the edges out in a $C_4\mathcal{T}$ invariant way to produce corner modes and TSC$_2$ topology. We analyze the topological invariants of 2d and 3d TSC$_2$s protected by $C_4\mathcal{T}$ symmetry and show that in both dimensions it is characterized by a $\mathbb{Z}_2$ topological invariant. Furthermore, we find that these topological invariants can be reduced to simple forms that depend only on the normal-state Fermi surfaces and the properties of the pairing, when in the weak-pairing limit. 

Besides focusing on mean-field BdG Hamiltonians, we show that certain interactions can favor the spontaneous formation of both TSC$_2$ scenarios in a weak-pairing picture. For the first scenario, we start with a normal state formed by a two-band Dirac nodal structure that can be realized in solid state or cold atom systems~\cite{demler2017}. We consider adding a chemical potential to the four Dirac nodes since, from the point of view of energetics, the presence of Fermi surfaces (FS) is beneficial for superconductivity, as the density of states  is finite (as opposed to vanishing linearly for the 2d Dirac points). Remarkably, we show that for a normal state  having four ``doped Dirac points" in the presence of a finite-range attractive interaction,
 a SC state with $p_x$ or $p_y$ pairing symmetry appears \emph{spontaneously} through a low-temperature instability. Further, we show by Landau-Ginzburg free energy analysis that a $p_x+ip_y$-wave order parameter is favored. We show that these ingredients are sufficient to generate the TSC$_2$ phase for the first scenario. 
 
 For the realization of the second scenario we take two different approaches. First, we consider a metallic system with a conventional spin-degenerate FS, and subject it to two types of electronic interactions that favor $p$-wave pairing and $d$-wave pairing respectively. These interactions, and their relevance to experiments, have been extensively studied previously~\cite{Fu-berg, brydon-sau-das-sarma, kozii,wang-cho-hughes-fradkin,brydon-agterberg,ruhman,ruhman-savary-Lee-fu,Chubukov2008}. In particular, for the $p$-wave order, it has been recently proposed~\cite{kozii,wang-cho-hughes-fradkin,ruhman} that fluctuations in the vicinity of an inversion symmetry breaking ordered phase induce $p$-wave order. For the $d$-wave order, perhaps the simplest mechanism is through the antiferromagnetic exchange interaction in an itinerant fermion system~\cite{Chubukov2008}. 
We show that the combination of these interactions naturally leads to the coexistence of $p$-wave and $d$-wave order. Following a similar Landau-Ginzburg free energy analysis, we show that the coexistence state indeed has $p+id$-wave order which is the desired form for the TSC$_2$ state. Additionally, we show that by coupling a $d$-wave superconductor with a 2d TSC will naturally produce the $p+id$ state and TSC$_2$ topology through the proximity effect. {In particular, we show that a heterostructure between FeTe$_{0.55}$Se$_{0.45}$~\cite{fesete1,fesete2,fesete5} and a cuprate SC can potentially realize a \emph{high}-$T_c$ TSC$_2$ phase.}


\section{TSC$_2$ from mirror-symmetric Dirac semimetal}
\label{sec:mirror}

 \subsection{A lattice model for TSC$_2$}
\begin{figure}
\centering
\includegraphics[width=0.6\columnwidth]{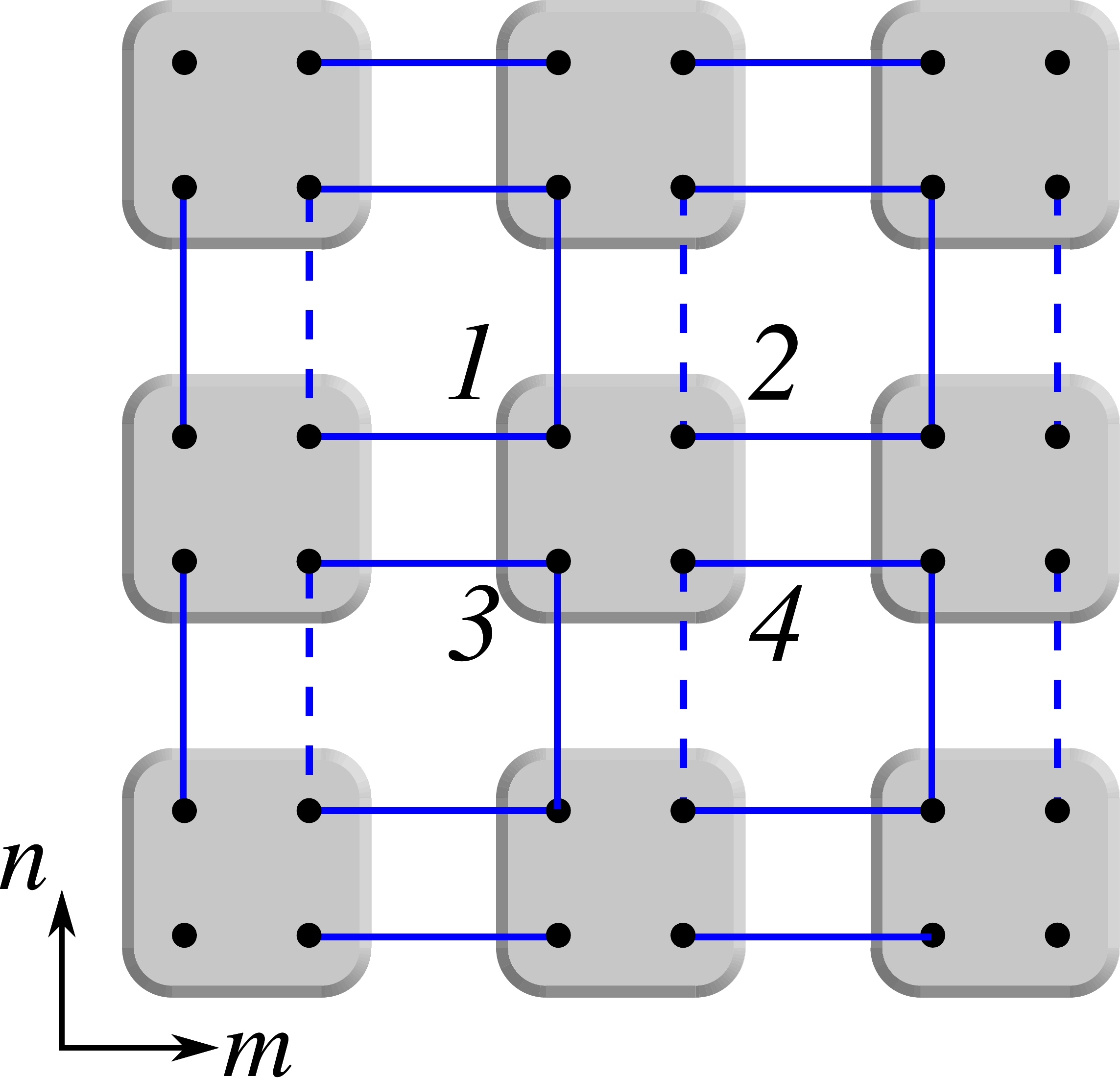}
\caption{Lattice representation of 2d second order topological superconductor Hamiltonian in a Majorana basis as in Eq.~\eqref{H0}. Each unit cell has four Majorana fermions represented by black dots as labelled. The tunneling strength $-2t$ is indicated by a solid line, while $2t$ is represented by a dashed line. As a result, there is a $\pi$-flux in each plaquette.}
\label{fig:plaq}
\end{figure}

We begin constructing a model for a 2d TSC$_2$ phase by close analogy with the quadrupole model in Ref.~\onlinecite{benalcazar2017}. That model is a tight-binding model on a square lattice with four complex fermion degrees of freedom per cell. If one simply replaces the four complex fermion orbitals by Majorana fermions, and replaces all of the hopping terms with Majorana tunneling terms, then one will have a model for a TSC$_2$ in a Majorana basis (see Fig. \ref{fig:plaq}). 
 The Hamiltonian in terms of Majorana operators is given by
\begin{align}
H=-2it\sum_{(m,n)}& \[\gamma^2_{m,n}\gamma^1_{m+1,n} +\gamma^4_{m,n}\gamma^3_{m+1,n} \right.\nonumber\\
&\left.-\gamma^2_{m,n}\gamma^4_{m,n+1} + \gamma^1_{m,n}\gamma^3_{m,n+1}\],
\label{H0}
\end{align}
where $(m,n)$ are the site coordinates.
The phases of the Majorana tunneling terms are tuned to have an effective $\pi$-flux per plaquette, and each plaquette is gapped. If we have boundaries of a sample then the edges are gapped, but have ``unpaired" Kitaev chains, and the corners harbor unpaired MBS (as shown in Fig. \ref{fig:plaq}). Thus, this is a natural model for a TSC$_2$ phase in 2d. 

Since two Majorana degrees of freedom represent one complex fermion degree of freedom, this model can physically describe a superconductor formed from a normal metallic state with two bands.
We can express the Majorana Hamiltonian in a complex fermion basis in terms of the hopping and pairing of electrons. To do this we combine the four Majorana operators per unit cell in pairs to form two complex fermions. There are several inequivalent ways one could choose to do this, and each one yields a different possible microscopic electronic realization of this TSC$_2$ phase. 

The choice of how to group the four Majorana modes per cell into two complex fermion modes essentially decides how the Hamiltonian splits into normal-state band structure and superconducting pairing gaps. Since our goal is to have the pairing terms generated as a low-temperature instability of the low-energy electrons, then it is desirable that we choose a microscopic realization such that the hopping terms lead to a gapless band structure, and the pairing terms describe its intrinsic superconducting tendency. Interestingly, this can be achieved by the following identification ($\uparrow$ and $\downarrow$ denote the two (pseudo) spin bands):
\begin{align}
c_{\uparrow,2m+1,n}=&(\gamma^1_{2m+1,n}+i\gamma^2_{2m+1,n})/\sqrt{2},\nonumber\\
c_{\downarrow,2m+1,n}=&(\gamma^3_{2m+1,n}+i\gamma^4_{2m+1,n})/\sqrt{2},\nonumber\\
c_{\uparrow,2m,n}=&(\gamma^3_{2m,n}+i\gamma^4_{2m,n})/\sqrt{2},\nonumber\\
c_{\downarrow,2m,n}=&(\gamma^1_{2m,n}+i\gamma^2_{2m,n})/\sqrt{2}.
\end{align}
 From this combination of Majorana operators the resulting four-band BdG Hamiltonian derived from Eq.~\eqref{H0} is given by 
$H= \int d{\bf k}\Psi^\dagger_{\bf k}
\mathcal{H}({\bf k})
\Psi_{\bf k}$
where $\Psi_{\bf k} = (c_{\bf k} , c_{-{\bf k}}^\dagger)^T$ and
 \begin{align}
\mathcal{H}({\bf k})=&t\cos k_x \sigma_x\tau_z + t\cos k_y\sigma_y \nonumber\\
&+ \Delta\sin k_x \sigma_x \tau_y + \Delta\sin k_y \sigma_x\tau_x,
 \label{eq:1} 
 \end{align}
 where we have allowed for two separate parameters $t$ and $\Delta$ (for which Eq. \eqref{H0} has $t=\Delta$), $\sigma_i$'s are Pauli matrices in the (pseudo) spin space,  $\tau_j$'s are Pauli matrices in Nambu space,  $\sigma_i\tau_j$ denotes their Kronecker product, and we have set the lattice constant $a_0=1.$
This BdG Hamiltonian has a particle-hole symmetry ${\mathcal C}=\tau_x$, such that $\mathcal{C} \mathcal{H}^T(-{\bf k})\mathcal{C}^{-1} = -\mathcal{H}({\bf k}).$

\begin{figure}
\centering
\includegraphics[width=\columnwidth]{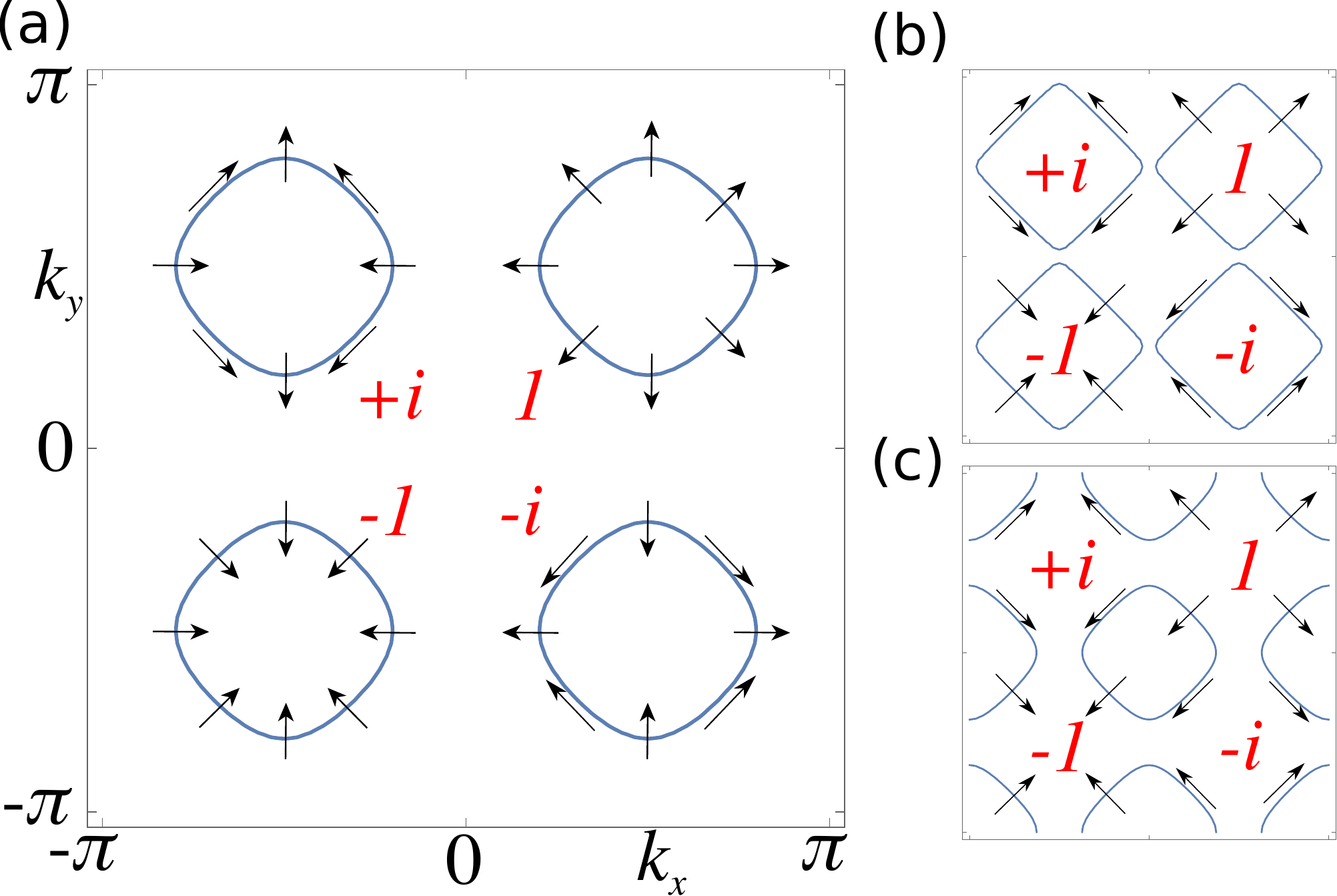}
\caption{The Fermi surface for the BdG Hamiltonian in Eq.~\eqref{eq:1} with $(t,\Delta)=(1,1/2)$, and (a) $\mu=0.90$, (b) $\mu=1.11$ and (c) $\mu=1.20$. The phase of the $p$-wave order parameter and the spin texture of  each pocket are indicated around each Fermi surface. Upon increasing the chemical potential $\mu$, the pockets will increase in sizes, and merge, and vanish at the high symmetry points in the BZ.  } 
\label{fig:2}
\end{figure}

Before discussing the full TSC$_2$ phase, let us consider just the normal state, two-band Bloch Hamiltonian with a chemical potential $\mu$ (whose effect on the topology we discuss later):
\begin{equation}
\mathcal{H}_{\rm N}({\bf k})=t\cos k_x \sigma_x +t\cos k_y \sigma_y-\mu,\label{eq:HN}
\end{equation} which has four gapless Dirac points, when $\mu=0,$ located at $(k_x, k_y)=(\pm \pi/2, \pm \pi/2).$ For finite $\mu,$ the system develops Fermi surfaces centered around each of the Dirac points, and for large values of $\mu>t$ the system will undergo a Lifshitz transition eventually leading to vanishing Fermi surfaces when $\mu>\sqrt{2}t$, i.e., when Fermi level lies outside bandwidth. This system has mirror symmetries $M_x=\mathbb{I}$ and $M_y=\mathbb{I}$ satisfying 
\be
\label{mirror}
M_{x,y} \mathcal{H}_{\rm N}({\bf k})M_{x,y}^{-1} = \mathcal{H}_{\rm N}(\hat m_{x,y}{\bf k}),
\ee where, e.g., $\hat m_x(k_x, k_y) = (-k_x, k_y)$. We note that  these operators obey $[M_x, M_y]=0$, and thus these mirror symmetries do not support higher-order topology~\cite{benalcazar2017}. Let us focus on the range $0<\mu<\sqrt{2}t$, for which there are {closed or open} Fermi surfaces centered at $(k_x, k_y)=(\pm \pi/2, \pm \pi/2)$, which we show in Fig.\ \ref{fig:2}. 
Each of the Fermi surfaces has a (pseudo)spin texture (see Fig.\ \ref{fig:2}), and crucially, as far as superconductivity is concerned, the portions of the FS's with opposite momenta always occur with the \emph{same} (pseudo) spin texture. This means that, as a weak-coupling instability, \emph{only} (pseudo) spin triplet, odd-parity (e.g., $p$-wave) pairing can occur, which is exactly what is required by nontrivial second-order topology according to Eq. \eqref{eq:1}.

Now let us tune back to $\mu=0$ so that we only have the four Dirac points, and consider the addition of the $p_x+ip_y$ pairing terms in Eq. \eqref{eq:1}. The superconducting gaps at the four Dirac points have a circulating phase structure as one moves from a Dirac point in one quadrant to another, with phases of $0, \pi/2, \pi,$ and $3\pi/2$ respectively as shown in Fig. \ref{fig:2}. These pairing terms \emph{break} the mirror symmetries of the normal state, however we can define a new set of mirror symmetries 
{
\be
\mathcal{M}_x= \sigma_y\tau_y, ~\mathcal{M}_y= \sigma_y\tau_x,
\ee
}
 such that $\mathcal{M}_{x,y} \mathcal{H}({\bf k})\mathcal{M}_{x,y}^{-1} = \mathcal{H}(\hat m_{x,y}{\bf k}).$ Crucially, these operators satisfy $\{\mathcal{M}_x,\mathcal{M}_y\}=0$ and can support gapped Wilson loop spectra and higher-order topology. Indeed, if one calculates the nested Wilson loops one finds that this system is in a non-trivial TSC$_2$ phase analogous to the quadrupole insulator, but with unpaired MBS on the corners of the sample instead of complex fermions. In fact, this is immediately manifest since we constructed our BdG Hamiltonian from a higher-order TSC$_2$ in the Majorana basis in Eq.~\eqref{H0}. To generate these new, non-commuting mirror reflections, the superconducting gap has to transform nontrivially under mirror symmetries in \emph{both} directions seperately. Thus, gapped superconductors with other possible pairing symmetries, such as $s$-wave, $p_x$-wave, or $p_y$-wave, will not generate mirror-protected second-order topology starting from this normal state Hamiltonian.

{We can add various perturbations to the Hamiltonian \eqref{eq:1}. Specifically, we consider four types of terms: $\mathcal{H}_1=-\mu\tau_z,~ \mathcal{H}_2=m\sigma_z\tau_z,~
\mathcal{H}_3= b_x\sigma_x\tau_z,~\mathcal{H}_4=b_y\sigma_y.$ The effect of $\mathcal{H}_2$ is to open gaps in the normal-state Dirac nodes, and it competes with the superconducting gap.  $\mathcal{H}_{3,4}$ shift the Dirac nodes in a mirror-symmetric fashion in the $k_x$ and $k_y$ directions respectively.
We find that, when individually turned on and tuned, 
 the conditions to remain in the topological phase are
\be
|\mu|<\sqrt{2}|t|,~|m|<\sqrt{2}|\Delta|,~|b_x|<|t|,~|b_y|<|t|.
\label{3}
\ee
This condition can be understood in two ways. In a momentum space picture, the critical values for $m, b_{x,y},$ and $\mu$ correspond to either gapping out ($m$) the Dirac nodes, shifting and annihilating ($b_{x,y}$) them, or shifting the chemical potential $\mu$ out of the bandwidth. In a real space picture in the Majorana basis shown in Fig.\ \ref{H0}, these terms correspond to onsite coupling terms between the Majorana modes. When the onsite couplings become larger than the inter-cell couplings, the system transitions into a trivial phase.

Interestingly, we note that the chemical potential $\mathcal{H}_1$ term actually breaks the mirror symmetries ${\mathcal{M}}_{x,y}$! However,  the higher order topological phase is still robust, as the bulk topological invariant is also protected by particle-hole symmetry $\mathcal{C}.$  The invariant that characterizes the higher order superconductor is the mathematical analog of the quadrupole moment $q_{xy}$ which is defined in the analogous insulator system as 
\begin{eqnarray}\begin{aligned}
\label{eq:def_qxy}
q_{xy} \equiv Q_{\text{Cor}} - p_x - p_y,
\end{aligned}\end{eqnarray}
for a square lattice with edges and corners. For superconductors we interpret $Q_{\text{Cor}}$ is the parity of the number of Majorana bound states, and $p_x$ ($p_y$) are the \textit{edge Berry phases} in units of $2\pi$ for the edges parallel to $\hat{x}$ ($\hat{y}$) ($p_{x/y}$ are interpreted as the edge polarizations in the context of quadruple insulator\cite{benalcazar2017}). Since particle hole symmetry is a local symmetry that flips the sign of the charge, it quantizes both $Q_{\text{Cor}}$ and $p_{x,y}$ to integer or half-integer values in the insulator case. As a result of Eq.~\eqref{eq:def_qxy}, $q_{xy}$ is also quantized, and the higher order topological phase is robust in the presence of particle hole symmetry. The model we consider is mathematically identical to the insulator system and hence has a quantized topological invariant protected by $\mathcal{C}$ alone.

However, to illustrate that this topology is a bulk property, it is desirable to calculate $q_{xy}$ in terms of bulk quantities, for example, in a periodic system with no edges or corners where Eq.~\eqref{eq:def_qxy} is not applicable. As shown in Ref.~\onlinecite{benalcazar2017}, this can be done via the nested Wilson loop which is quantized by the mirror symmetries, and is not quantized by particle hole symmetry alone. We discuss the role of mirror symmetries in more detail  in Sec.~\ref{sec:A more general condition for TSC2}, and leave the identification of a purely bulk expression for the invariant in the presence of particle hole symmetry alone to future work.

\subsection{A more general condition for TSC$_2$}
\label{sec:A more general condition for TSC2}
 For many examples of TSCs in the weak-coupling limit, it can be shown that the topological characterization of the system is \emph{completely} determined by the properties of the normal state FS and SC order parameter~\cite{qi-hughes-zhang}. A topological invariant that can be determined from the low-energy physics alone serves as a useful tool for identifying and searching for TSC states in real materials.
  However, we find that the topological invariant of the mirror-protected TSC$_2$ cannot be reduced to the low-energy quantities near the FS, at least in the linearized limit. Heuristically, determining the topology through the nested Wilson loops relies on the properties of the Wannier bands~\cite{benalcazar2017}, not the energy bands. Therefore, low-\emph{energy} physics near the Fermi energy does not necessarily completely capture the topology, even in the weak-coupling limit. {Indeed, one can consider a case where there are four, mirror-related Dirac nodes, but which do not arise from a single pair of bands. They could arise from two pairs of bands, each with two Dirac nodes, or four pairs of bands, each with a single Dirac node. After turning on a (necessarily inter-band) $p_x+ip_y$ SC order parameter, the low-energy theory is identical to our model; their differences are encoded in how the Dirac points are connected at high energies. We found that these models where all four normal-state Dirac nodes are not connected at high energies (i.e., when they do not all arise from the same pair of bands) generally do not support higher order topology in the presence of $p_x+ip_y$ pairing.} We will see in our second TSC$_2$ scenario that for some symmetry classes, i.e., at least for the $C_4\mathcal{T}$ class, we \emph{can} find low-energy topological invariants that describe the higher-order topology. It may also be possible to circumvent this problem by considering other symmetry classes beyond the ones studied here, or by keeping track of the low-energy physics of both the bulk and the boundaries, which may be sufficient to capture the Wannier band topology. We will leave such considerations to future work.
  
{ Despite this difficulty for the mirror-symmetric TSC$_2$, one can prove the following \emph{sufficient} condition for a TSC$_2$ phase based purely on low-energy considerations: for a two-band doped Dirac semimetal that is mirror symmetric (satisfying  Eq.\ \eqref{mirror}) with $H_{\rm N}'=\int d{\bf k}c^\dagger({\bf k})\mathcal{H}_{\rm N}'({\bf k}) c({\rm k})$ where
\be
{\mathcal{H}}_{\rm N}'({\bf k}) = f_1(k_x, k_y) \sigma_x + f_2(k_x, k_y)\sigma_y - \mu
\ee
with $\mu$ inside the bands, and four Dirac nodes at $(\pm k_x^{\ast}, \pm k^{\ast}_y)$ with $k_x^{\ast}, k_y^{\ast}\neq 0$, a fully gapped $p_x+ip_y$-wave SC order realizes a TSC$_2$ phase. By mirror symmetry \eqref{mirror} $f_1$ and $f_2$ are real, even functions of both $k_x$ and $k_y,$ and we assume they have a simultaneous zero at a generic point in the Brillouin zone;  mirror symmetry implies zeros at four points  $(\pm k_x^{\ast}, \pm k^{\ast}_y)$. At any chemical potential $\mu$ inside the bands, the (pseudo) spin texture of the fermions near the Fermi level lies in the $x,y$ (pseudo) spin plane, and the spin orientations for  $\bf k$ and $-\bf k$ are the {\it same} at the Fermi level.
One can straightforwardly show that the pairing order 
\be
H_{p}'=\int d{\bf k} \Delta({\bf k}) c^\dagger({\bf k}) \sigma_x [c^\dagger(-{\bf k})]^T+h.c.,
\ee gaps out $\pm\bf k$ points with the same spin orientation in our case. Further, we focus on fully-gapped, $p$-wave pairing where 
\be
\Delta({\bf k}) = g_1(k_x,k_y) +i g_2(k_x,k_y),
\ee
where $g_1({\bf k})$ and $g_2({\bf k})$ are real, odd functions and vanish only at time-reversal invariant momentum points $k_{x,y}=0$ or $\pi$. (We assume that aside from these symmetry enforced nodes $g_{1,2}$ do not have any other accidental nodes.)
The BdG Hamiltonian of this SC state is given by
\begin{align}
\label{eq:bdg'}
\mathcal{H}'({\bf k})=&f_1({\bf k})\sigma_x\tau_z + f_2({\bf k}) \sigma_y -\mu\tau_z \nonumber\\
&+ g_1({\bf k})\sigma_x\tau_x + g_2({\bf k})\sigma_x\tau_y.
\end{align}}

{We now argue that this gapped phase described by Eq.\ \eqref{eq:bdg'} is in a TSC$_2$ phase. Let us focus on the case with $\mu=0$ first. We can think of a process to obtain this generic Hamiltonian $\mathcal{H}'$ by deforming Eq.~\eqref{eq:1} while maintaining, e.g, the mirror symmetries. Namely, we define $\mathcal{H}'({\bf k}, \alpha)$ in the same form as \eqref{eq:bdg'} with $\alpha\in (0,1)$  where $f_1({\bf k}, 0)\equiv\cos k_x$ and $f_1({\bf k}, 1)\equiv f_1({\bf k})$ and similarly define $f_2({\bf k}, \alpha)$ and $g_{1,2}({\bf k}, \alpha)$.
During the deformation process the Dirac points given by the normal-state part of the BdG Hamiltonian do not collapse and annihilate, and the bulk gap remains open since the equal-spin pairing term always gaps out the Dirac points. 
While this shows we can continuously connect these Hamiltonians without the bulk gap closing, we also need to show that the corner Majorana modes in Eq.~\eqref{eq:1} do not disappear due to a Wannier transition either, i.e., a bulk-driven transition of the edge Hamiltonian.~\cite{klich,benalcazar2017,benalcazarPRB} In Appendix \ref{app:a} we prove that such a Wannier transition does not occur \emph{as long as} the normal state Dirac points do not annihilate. If we then include a chemical potential $\mu$ in Eq.\ \eqref{eq:bdg'} it is straightforward to show that, for sufficiently small $\mu,$ neither the bulk nor edge spectrum undergo a transition. Therefore we have shown that Eq.\ \eqref{eq:bdg'} realizes a TSC$_2$.}

As an explicit example of this condition we can consider the Hamiltonian  \begin{align}
\mathcal{H}_{b}({\bf k})=&(b_x+t\cos k_x) \sigma_x\tau_z + t\cos k_y\sigma_y \nonumber\\
&+ \Delta\sin k_x \sigma_x \tau_y + \Delta\sin k_y \sigma_x\tau_x,
 \label{eq:11},
 \end{align}
which we have already found is a TSC$_2$ for $|b_x|<|t|$. We note that precisely within this range, the BdG Hamiltonian describes a $p_x+ip_y$ superconductor with a normal state with four mirror-symmetric Dirac points. At $b_x=-t$ the normal-state Dirac points are maximally shifted and annihilate on the $k_x=0$ axis. Interestingly, at this point the system goes through a Wannier transition while the bulk gap remains open. To see this, we can calculate the effective Hamiltonian for the top and bottom edges (open boundaries in the $y$-direction). {From the second and fourth term in Eq.\ \eqref{eq:11} the wavefunction of the edge states satisfy $\sigma_z\tau_x\Psi^{\rm b/t}(k_x, y)=\pm \Psi^{\rm b/t}(k_x,y)$, where b/t denotes bottom and top edge respectively. The edge Hamiltonians are given by
\be
\mathcal{H}^{\rm b/t}_{b}(k_x) = (b_x+t\cos k_x)\mu_x^{\rm b/t} - \Delta \sin k_x\mu_y^{\rm b/t}.
\ee
where $\mu_i^{\rm b}$ is the Pauli matrices in the subspace of ${|\Uparrow\rangle}_{\rm b} \equiv |\uparrow\rangle_{\sigma}\otimes |\rightarrow\rangle_{\tau} $ and ${|\Downarrow\rangle}_{\rm b} \equiv |\downarrow\rangle_{\sigma}\otimes |\leftarrow\rangle_{\tau} $, and $\mu_i^{\rm t}$ is the Pauli matrices in the subspace of ${|\Uparrow\rangle}_{\rm t} \equiv |\uparrow\rangle_{\sigma}\otimes |\leftarrow\rangle_{\tau}  $ and ${|\Downarrow\rangle}_{\rm t} \equiv |\downarrow\rangle_{\sigma}\otimes |\rightarrow\rangle_{\tau} $.}
One can straightforwardly verify that this edge Hamiltonian becomes gapless and transitions from topological to trivial at $b_x=-t$. From this example we see that the Wannier transition is tied to the fate of the normal-state Dirac points, and as long as the Dirac points do not annihilate the system generates TSC$_2$ topology with the $p_x+ip_y$ pairing.

   \subsection { Realization from electronic interactions}
The fact that our proposed superconducting Hamiltonian \eqref{eq:1} has a gapless normal-state band structure indicates that the required superconducting gap can potentially be \emph{intrinsically} induced from electronic interactions. From an energetic perspective, the presence of Fermi surfaces at a finite $\mu$ greatly enhances the pairing instability so we will consider  a normal state of ``doped" Dirac points with Fermi surfaces at finite $\mu.$ We have shown above that doing so does not change the topology of the superconducting state in which we are interested. To be specific, we consider the same nearest-neighbor tightbinding Hamiltonian as in Eq.~\eqref{eq:HN} with a finite $\mu$ 
with $0<\mu<t$ (The situation with four closed pockets around each Dirac points, shown in Fig. \ref{fig:2}).  There are four Fermi pockets centered at $(k_x, k_y)=(\pm \pi/2, \pm \pi/2)$, and the regions of the FS's with opposite momenta always occur with the \emph{same} (pseudo) spin texture, hence naturally leading to triplet, odd-parity pairing instead of, e.g., singlet $s$-wave pairing. 
Unlike topology, the superconducting critical temperature $T_c$, as well as the exact form and magnitude of the superconducting gap $\Delta$ are not universal properties, and depend on microscopic details such as the band dispersion and the structure of the electronic interactions. However, the remarkable feature that odd-parity ($p$-wave) pairing is expected to be dominant over $s$-wave pairing for our normal state system is an encouraging sign  for its realization. 

We now move on to study a concrete pairing mechanism. The sign-changing structure of the $p$-wave order parameter in $\bf k$-space places restrictions on the required $\bf k$-space structure of the electronic interactions. Indeed, momentum-independent electron-phonon interactions do not induce pairing in the $p$-wave channel at weak coupling, because, within the ladder approximation~\footnote{Quantum corrections beyond the ladder approximation can induce $p$-wave instabilities; see, e.g., W. Kohn and J. M. Luttinger, Phys. Rev. {\bf 15}, 525 (1965), and A. V. Chubukov and S. A. Kivelson, Phys. Rev. B {\bf 96}, 174514 (2017).}, the contribution to the pairing susceptibility from FS regions with positive and negative pairing gaps cancels out. For our purposes, we consider a density-density interaction given by the effective action 
\begin{align}
\!\!\!\!S_{\rm int}= - \int dk dq D(q) c_\alpha^\dagger(k)c_\alpha(k) c_{\beta}^\dagger(k+q)c_{\beta}(k+q),
\end{align}
 where $\alpha,\beta$ are (pseudo) spin indices and are summed over,  $k \equiv (\omega_m, {\bf k})$, $q \equiv (\Omega_m, {\bf q})$, and $\omega_m, \Omega_m$ are Matsubara frequecies. $D(q)$ can be thought as the propagator of a collective mode, and for simplicity we take an Ornstein-Zernike form 
\be
D(\Omega,{\bf q})=1/(\Omega^2+c^2{\bf q}^2+c^2\xi^{-2}).
\ee
This propagator is peaked at zero momentum, and it can be realized physically by  fluctuations of an electronic nematic order,~\cite{lederer} or a soft optical phonon mode with a strong momentum dependence peaked at ${\bf q}=0$. For example, such a phonon mode has been proposed to play an important role in high-temperature superconductivity in monolayer FeSe on SrTiO${_3}$.~\cite{dunghai-1,dunghai-2}


We make three further simplifications. {First, we assume that the Fermi pockets in Fig.\ \ref{fig:2}(a) are circular.} Second, we take the weak-coupling limit and neglect all self-energies and vertex corrections. Third, we assume that the correlation length in units of the lattice constant $a_0$ satisfies
$1\ll\xi/a_0\ll t/\mu$, 
 such that the intra-pocket interaction can be treated as constant, and dominates over the inter-pocket ones. With these assumptions, the linearized gap equation for the $p$-wave order $\Delta_1$ between the two pockets centered at $\pm (k_x^\ast, k_y^\ast)$ is
\begin{align}
\Delta_{1} = \frac{\lambda_0-\lambda_2}{4}{N(0)}\Delta_1\ln\frac{\Lambda}{T}\int \frac{d\theta}{2\pi} \cos^2\frac{\theta}{2},
\end{align}
where we have defined $\lambda_0\equiv D(\Omega=0, {\bf q}=0)$, $\lambda_1\equiv D(\Omega=0, |{\bf q}|=\pi)$, and $\lambda_2 \equiv D(\Omega=0,{\bf q}=(\pi,\pi))$. We note that $\lambda_1$ dependent terms happen to cancel and not enter the equation. {The $\ln(\Lambda/T)$ factor corresponds to the standard Cooper instability, where $\Lambda$ is an ultraviolet cutoff and $T$ is the temperature.} $N(0)$ is the density of states at the Fermi level. Additionally, the angular integrand $\cos^2(\theta/2)$ obtains from the spin-texture on the FS's.

We can extract the superconducting critical temperature as 
\be
T_c= \Lambda \exp\[\frac{-8}{(\lambda_0-\lambda_2)N(0)}\].
\ee
By the spatial symmetry of our system, the analysis for the $p$-wave order $\Delta_2$ between the two pockets centered at $\pm (k_x^\ast, -k_y^\ast)$
 follows analogously, and the resulting $T_c$ is identical. The interplay between the $\Delta_{1,2}$ orders can be addressed within a Ginzburg-Landau free energy formalism:
\begin{align}
F=& \alpha(|\Delta_1|^2+|\Delta_2|^2) +\beta (|\Delta_1|^4+|\Delta_2|^4)\\
& +4\beta'|\Delta_1|^2|\Delta_2|^2 + \beta''\[\Delta_1^2(\Delta_2^*)^2 + \Delta_2^2 (\Delta_1^*)^2\].\nonumber
\end{align}
Whether, and how, the $\Delta_1$ and $\Delta_2$ order parameters coexist is determined by the quartic terms. Since in our case $\Delta_1$ and $\Delta_2$ couple to different pockets, their competition effects (which are captured by the $\beta'$ term) are small, and $\Delta_1$ and $\Delta_2$ coexist in the ground state. From the $\beta''$ term, no matter how small, the relative phase between $\Delta_{1,2}$ is fixed to be $\pm \pi/2$.~\cite{maiti-chubukov-2013,wang-chubukov-2014,congjun-trsb} It is straightforward to check the phase of the SC gap on each pocket to find that such a coexistence state is indeed a $p_x+ip_y$ SC state.
 Therefore, we have shown that via a simple pairing mechanism, a two-dimensional Dirac system \emph{precisely} realizes the second-order topological superconductivity spontaneously.

%
%

\section{TSC$_2$ from a $C_4\mathcal{T}$ symmetric superconductor}

\subsection{$p+id$ pairing symmetry}

In this section we discuss another type  of TSC$_2$ phase in both 2d and 3d characterized by a combined symmetry of $C_4$ spatial rotation and time-reversal $\mathcal{T}.$ We consider the following Hamiltonian
\begin{align}\label{BdG_c4tA}
H =& \int d{\bf k}\[ c^\dagger({\bf k}) \(\frac{{\bf k}^2}{2m}-\mu\) c({\bf k}) \right.\nonumber\\
& \left.+ \Delta_pc^T({\bf k}) ({\bf k}\cdot \vec \sigma) i\sigma^y c({\bf -k}) \right.\nonumber\\
& \left. + i  \Delta_{d} c^T({\bf k})(k_x^2 - k_y^2) i\sigma^y c({\bf -k}) + h.c.\],
\end{align}
which can be used in both 2d and 3d.
 The first term describes an ordinary spin-degenerate Fermi surface, and the second term corresponds to a time-reversal invariant $p$-wave pairing, commonly denoted as $(p+ip)/(p-ip)$ order in 2d, or the analog of the superfluid $^3$He-$B$ phase in 3d\cite{schnyder2010,raghu}. The first two terms have time reversal symmetry $\mathcal{T},$ as well as a particle-hole symmetry $\mathcal{C}$. The third term is a $d$-wave pairing term, which is odd under a $C_4$ lattice rotation, with a relative phase of $\pi/2$ with respect to the $p$-wave order. For convenience, we take this phase difference into account by treating the $d$-wave order parameter as imaginary, and we denote the pairing symmetry of this SC state as $p+id$. Owing to the imaginary $d$-wave pairing term, such a superconducting state breaks both time-reversal symmetry $\mathcal{T}$ and $C_4$ rotational symmetry, but is invariant under the combined $C_4\mathcal{T}$ operation.
 
 Such a SC model supports chiral Majorana modes on the hinges of a sample in 3d, or MBS on the corners of a sample in 2d. We can understand the origin of these topological modes in a simple picture. For example, in 3d, the $p$-wave superconducting order by itself realizes topological superconductivity in class DIII, which supports gapless Majorana cones on all surfaces. The addition of the bulk $d$-wave order parameter  gaps out these surface Majorana cones, as its relative $\pi/2$ phase with the $p$-wave order breaks $\mathcal{T}$. Since the $d$-wave order parameter changes sign under a $C_4$ rotation in the $xy$-plane, the Majorana masses for the neighboring side surfaces (parallel to the $z$-axis), say $xz$ and $yz$ surfaces, are opposite. Therefore, the hinges separating these surfaces can be viewed as mass domain walls for the surface Majorana fermions, and therefore they localize chiral Majorana modes. 
 This argument holds similarly in 2d to generate single MBS at corners from mass domain walls of the initially-gapless helical Majorana modes on the edges.
 We illustrate these MBS in Fig.\ \ref{fig:p_d}.
 
 \begin{figure}
\centering
\includegraphics[width=\columnwidth]{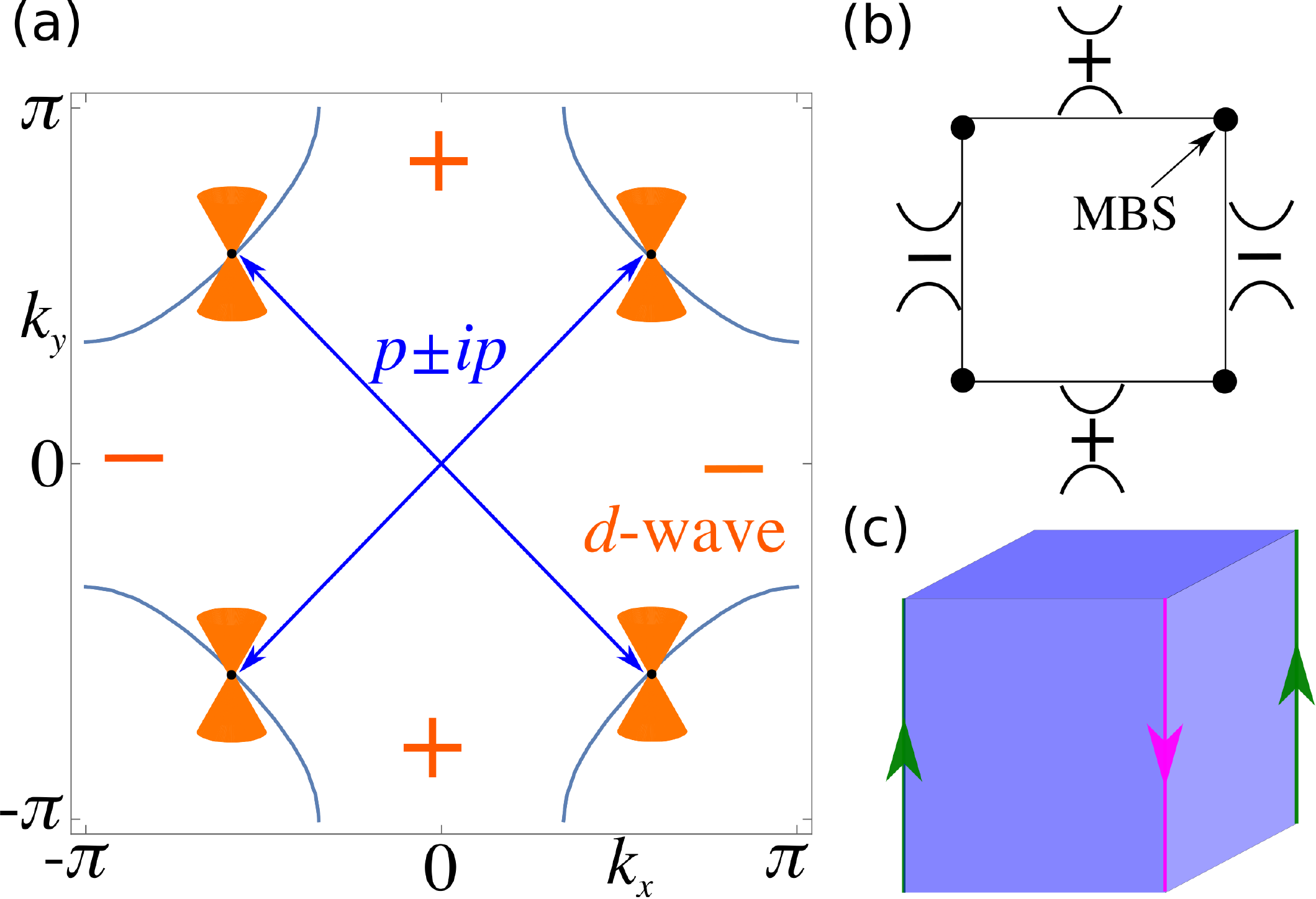}
\caption{(a) The normal state band structure of a cuprate. With the $d$-wave order nodes are generated, as indicated by the four Dirac cones along the nodal directions. The phase of the $d$-wave order parameter is indicated by the sign structure in orange. In the presence of $p$-wave order, the Dirac cones in the nodal directions are coupled, as shown in blue, which results in a completely gapped superconducting order. (b) A $\text{TSC}_2$ in 2D with gapped edges. The Majorana gaps have opposite signs on neighboring edges such that there is a MBS residing at the corner. (c) A $\text{TSC}_2$ in 3D with gapped surfaces, and gapless Majorana modes along the hinges. The colors red and green indicate that the neighboring hinges have different chiralities.}
\label{fig:p_d}
\end{figure}
 
For the sake of completeness,  we note that there is another set of $C_4$ and $\mathcal{T}$ broken, but $C_4\mathcal{T}$ invariant, terms allowed in the superconducting system. Such terms are given by, for example, 
 \be
 \int d{\bf k} c^\dagger ({\bf k}) (k_x^2-k_y^2) \sigma_i c({\bf k}),~ {i=x,y,z},
 \ee
  and represent  spin-nematic order that might be induced as a Pomerunchuk instability in the spin channel.\cite{wu-sun-fradkin-zhang,fischer-kim,rodrigo-pdw} These terms deform the Fermi surfaces in a spin-dependent way. However, we found that these terms do not fully gap the system. Choosing $i=x,y,$ or $z$, either the bulk becomes gapless ($i=x,y$), or the edges remain gapless ($i=z$). Therefore these perturbation terms, although allowed by symmetry, do not generate higher-order topology from our normal state Hamiltonian.

We also note that in 2d, this Hamiltonian \eqref{BdG_c4tA} has a similarity with the second-order TSC Hamiltonian  \eqref{eq:1} in the previous section.
The first and the third terms in Eq.~\eqref{BdG_c4tA} describe a regular $d$-wave SC, which has four quasi-particle Dirac nodes in the four diagonal directions in the BdG spectrum. Analogous to the discussion in the previous section, to construct a second-order TSC one needs to further gap out these Dirac points using a $p$-wave pairing.  It is straightforward to show that in this situation the $p$-wave order that can further gap out the Dirac nodes here is the $\mathcal{T}$-invariant  $p+ip/p-ip$  type with an overall phase difference $\pi/2$ with the $d$-wave order.
Such a superconductor is equivalent to the following lattice-regularized BdG Hamiltonian
 \begin{align}
 \label{BdG_c4t}
\mathcal{H}= &[(2-\cos k_x -\cos k_y)/m-\mu]\sigma_0\tau_z  \nonumber\\
&{- \Delta_p(\sin k_x \sigma_z\tau_x + \sin k_y \sigma_0 \tau_y )}\nonumber\\
&+ \Delta_d(\cos k_x - \cos k_y)\sigma_y\tau_x. 
\end{align} 
In this basis the time-reversal operator is given by $\mathcal{T}=i\sigma_yK,$ and the particle-hole operator is given by $\mathcal{C}=i\sigma_x\tau_yK$. It has a similar mathematical structure to what was proposed for a second-order topological insulator in a recent work Ref.~\onlinecite{schindler2017}. Here we show that it also has a natural interpretation as a TSC$_2$ model.

\subsection{Topological invariant of the $C_4\mathcal{T}$ symmetric superconductor}
We first focus on the 3d case. In the presence of time-reversal symmetry, topological superconductors in class DIII have a $\mathbb{Z}$ classification in 3d. This integer topological invariant can be computed from the bulk properties via a winding number $\nu$.\cite{schnyder}. 
We quickly review the key derivation of the winding number here.~\cite{schnyder, qi-hughes-zhang} 

The combination of particle-hole symmetry $\mathcal{C}$ and $\mathcal{T}$ gives rise to a chiral symmetry $\chi$, which ensures that the BdG Hamiltonian can be unitarily transformed into a block off-diagonal form. Let us take a $p$-wave topological superconductor as an example:
\begin{align}
\label{eq:BdG_p_no_d}
H= \int d{\bf k}\Psi^\dagger_{\bf k}
\begin{bmatrix}
\epsilon_{\bf k} & \Delta_p({\bf k}\cdot \vec \sigma)(i\sigma^y) \\
\Delta_p(-i\sigma^y)({\bf k}\cdot \vec \sigma) & -\epsilon_{{\bf k}}
\end{bmatrix}
\Psi_{\bf k},
\end{align}
where $\epsilon_{\bf k}={\bf k}^2/2m-\mu$ and $\Psi_{\bf k} = (c_{\bf k} , c_{-{\bf k}}^\dagger)^T$. Owing to the chiral symmetry $\chi$, we can rewrite the Hamiltonian as 
\begin{align}
\label{eq:BdG_p_no_d2}
H= \frac{1}{2}\int d{\bf k}\tilde{\Psi}^\dagger_{\bf k}
\begin{bmatrix}
 & Q_{\bf k} \\
Q_{\bf k}^\dagger & 
\end{bmatrix}
\tilde{\Psi}_{\bf k},
\end{align}
where 
\begin{align}
\tilde{\Psi}_{\bf k} \equiv &(c_{\bf k}+\sigma^yc_{-{\bf k}}^\dagger , c_{\bf k}-\sigma^yc_{-{\bf k}}^\dagger)^T, \nonumber\\
 Q_{\bf k}\equiv&\epsilon_{\bf k} - i\Delta_p{\bf k}\cdot \vec \sigma.
 \end{align} 
 
 In general $Q_{\bf k}$ is an $N\times N$ matrix, and can be decomposed via a singular value decomposition as $Q_{\bf k} = U^\dagger_{\bf k}D_{\bf k}V_{\bf k}$, where $U_{\bf k}$ and $V_{\bf k}$ are unitary. $D_{\bf k}$ is a diagonal matrix that consists of the positive eigenvalues of $H$. We can adiabatically tune the matrix $D_{\bf k}$ to the identity matrix $\mathbb{I}$ such that $Q_{\bf k}$ is deformed to a unitary matrix $q_{\bf k}\equiv U^\dagger_{\bf k}V_{\bf k}\in U(N)$. The topological invariant is the integer winding number $\nu$ of $q_{\bf k}$ defined as 
\be
\label{eq:winding1}
\nu = \frac{1}{24\pi^2}\int d{\bf k}\epsilon^{ijk}\Tr\left[ q_{\bf k}^\dagger \partial_i q_{\bf k} q_{\bf k}^\dagger \partial_j q_{\bf k} q_{\bf k}^\dagger \partial_k q_{\bf k} \right],
\ee which captures the homotopy class $\pi_3(U(N))=\mathbb{Z}.$ 

By construction, it can be proven that this integer topological invariant corresponds to the number of stable, gapless Majorana cones on the surface of a class DIII TSC. Moreover, for a weak-coupling superconductor in which the SC gap is only significant near the Fermi surface, this winding number can be  conveniently expressed in terms of the low-energy properties at the Fermi surface. Specifically, it was obtained in Ref.\ \onlinecite{qi-hughes-zhang} that
\begin{align}
\label{eq:winding2}
\nu = \frac{1}{2} \sum_{i} \sgn(\Delta_i) C_i,
\end{align}
where $i$ labels each non-spin-degenerate, $\mathcal{T}$-invariant FS in the normal-state, $\Delta_i$ is the sign of the SC gap on the $i$-th FS (time-reversal symmetry ensures all SC gaps can be made real), and the Chern number $C_i$ is the (quantized) net flux of the Berry phase gauge field piercing each FS. For a single-band, spin-$\frac{1}{2}$ system, the requirement for a TSC is simply that the signs of the superconducting pairing on the two spin-split FS's are opposite. 


Since our construction of a 3d $\text{TSC}_2$ here is closely tied to class DIII TSC, the question now is whether it also has a $\mathbb{Z}$ classification in 3d. For our $C_4\mathcal{T}$ symmetric BdG Hamiltonian  Eq.~\eqref{BdG_c4tA}, since $\mathcal{T}$ is broken, there is no conventional chiral symmetry, and one generally cannot transform its BdG Hamiltonian to an off-diagonal form. However, due to the $C_4\mathcal{T}$ symmetry, all the $\mathcal{T}$-breaking terms in the Hamiltonian are also odd under $C_4$ rotation. After a unitary transformation, it is then possible to rearrange the Hamiltonian such that the $C_4$ and $\mathcal{T}$-symmetric terms of the Hamiltonian are in the off-diagonal block, while the $C_4$-odd part is in the diagonal block. As an example, for Eq.~\eqref{BdG_c4tA}, we obtain
\begin{align}
\label{eq:BdG_p_d}
H= \frac{1}{2}\int d{\bf k}\tilde{\Psi}^\dagger_{\bf k}
\begin{bmatrix}
-\Delta_d(k_x^2-k_y^2) & Q_{\bf k} \\
Q_{\bf k}^\dagger & \Delta_d(k_x^2-k_y^2)
\end{bmatrix}
\tilde{\Psi}_{\bf k}.
\end{align}
Furthermore, the $\mathcal{T}$-invariant $p$-wave part of the Hamiltonian is already fully gapped in the bulk, which ensures the BDG Hamiltonian is fully gapped for a generic $d$-wave order.

Just like for  the class DIII TSC in 3d, the fully-gapped, chiral symmetric part of the Hamiltonian $Q_{\bf k}$ can be characterized by $\pi_3(U(N))=\mathbb{Z}$, i.e., the winding number defined in Eq.~\eqref{eq:winding1}. Specifically,  one can extract just the block-off diagonal ($\mathcal{T}$-symmetric) part of the Hamiltonian, which is fully gapped on its own, and calculate its winding number. However, this procedure raises the question of whether the winding number defined for  \emph{only a part} of the Hamiltonian actually has any physical meaning. To this end, one needs to verify whether it is tied to any topological properties. For example, one can check if it is necessary to close a gap in the bulk spectrum to generate a change in $\nu,$ and one should determine the relation between $\nu$ and any boundary/hinge modes that are stable against symmetry allowed perturbations. We will now illustrate both of these properties.

To gain some intuition, we first show that in a two-band, weak-coupling superconductor, the change from $\nu=1$ to $\nu=0$ in the presence of the $C_4\mathcal{T}$ symmetry, necessarily involves a bulk gap closing via \emph{Weyl points} in the BdG quasiparticle spectrum. Ref.\ \onlinecite{wang-fu-2017} showed that for a $\mathcal{T}$-invariant weak-coupling superconductor, {generally in the presence of discrete lattice symmetries}, the transition from $\nu=1$ to $\nu=0$ is induced by the creation and annihilation of pairs of nodal lines on one of the spin-split FS's. We illustrate this process in Fig.\ \ref{fig:nodalline}(a). We have used ellipsoidal FS's to illustrate the lack of full rotational symmetry, but the scenario applies to generic FS's in lattice systems. From the formula in Eq.\ \eqref{eq:winding2}, the leftmost configuration in Fig.\ \ref{fig:nodalline}(a) is a trivial SC phase and the rightmost configuration is a TSC phase. There is an intermediate gapless state separating these two gapped phases where nodal lines separate ``puddles" of positive and negative superconducting gap function on a FS. 

Now we can evaluate if this critical transition region is destroyed when we break $\mathcal{T},$ but preserve $C_4\mathcal{T}.$
For our TSC$_2$ model,  the presence of the imaginary $d$-wave gap generates this symmetry breaking. In the presence of this term we find that most of the pieces of the nodal lines in the $\mathcal{T}$-symmetric sector are gapped. However, the $d$-wave gap here necessarily 
has nodes that are related by $C_4$ rotations. 
Where the $d$-wave nodes intersect the nodal lines there will be Weyl nodes in the BdG quasiparticle spectrum~\cite{fu-weyl-majorana}. While, in general, the nodal lines in the $\mathcal{T}$-invariant limit may have more complicated geometry, it is straightforward to see that the Weyl-nodal intermediate state is \emph{unavoidable} through this transition. As Weyl points cannot be gapped on their own, and they are related by a $C_4$ rotation, such a gapless intermediate state is stable. Therefore, the transition between a $C_4\mathcal{T}$ symmetric second-order TSC and a trivial SC occurs via an interesting Weyl-nodal SC phase and is captured by the changing of the winding number $\nu$ of just the gapped, $\mathcal{T}$-symmetric part of the Hamiltonian.

 \begin{figure}
\centering
\includegraphics[width=\columnwidth]{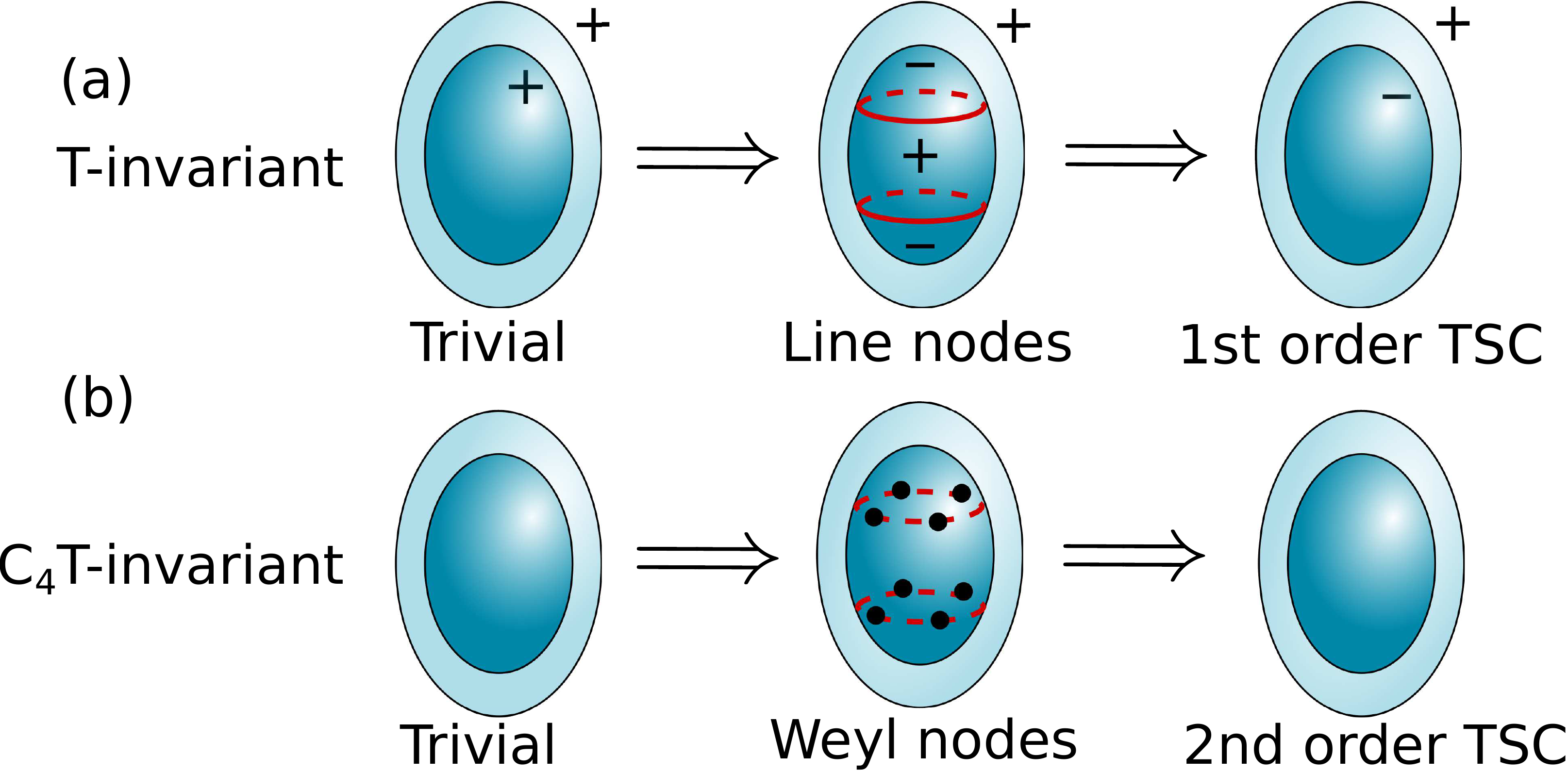}
\caption{Topological phase transitions for (a) $\mathcal{T}$-invariant TSC and (b) $C_4\mathcal{T}$-invariant 2nd order TSC. In the trivial phase of a $\mathcal{T}$-invariant TSC,  the winding number is $\nu=0$ as the SC gaps on the two FS's have the same sign. A topological transition occurs when a pair of nodal lines are created on one of the FS's, and the SC gap changes from positive to negative on this FS as the nodal lines nucleate, sweep over the FS, and eventually annihilate. The TSC enters the topological phase with $\nu=1$, after the nodal lines are annihilated. In the presence of $C_4\mathcal{T}$, however, a $d$-wave order can gap most out the nodal lines except at eight Weyl points. When the Weyl points are annihilated, a 2nd order TSC is formed. }\label{fig:nodalline}
\end{figure}

Interestingly,  the same argument \emph{does not} hold for two copies of TSC$_2$, i.e., when $\nu=2.$  The issue is that with a four band, i.e., four (spin-split) FS model, it is possible to have {\it fully gapped} $d$-wave order~\cite{chubukov-vafek-fernandes} such that the Weyl nodes will not be created. Labeling the two copies of the TSC$_2$ phase with $\mu_z =\pm 1$, such a pairing term can be written as
\be
H_d = i  \int d{\bf k} c^T({\bf k}) [(k_x^2 - k^2_y) \mu_z + (k_xk_y) \mu_x] i\sigma^yc(-{\bf k})+ h.c.,
\label{dxy}
\ee  
which is a non-commuting combination between $d_{x^2-y^2}$-wave and $d_{xy}$-wave order. This imaginary $d$-wave pairing term completely gaps out the Fermi surface, including the would-be nodal lines in the $\mathcal{T}$-symmetric sector during a transition from $\nu=2$ to $\nu=0$. Thus, this transition can occur without a gap closing, hence $\nu=2$ and $\nu=0$ belong to the {\it same} phase. This result indicates that the topological invariant is a $\mathbb{Z}_2$ quantity given by $P\equiv(-1)^\nu$.

The identification of $P\equiv(-1)^\nu$ as a bulk topological invariant can also be established via the stability of the hinge modes. As we discussed, for the $\mathcal{T}$-invariant system the winding number $\nu\in \mathbb{Z}$ corresponds to the number of stable, surface Majorana cones. When the  imaginary $d$-wave order parameter is turned on it gaps out the Majorana cones on surfaces parallel to the $z$-axis, and induces chiral Majorana modes at the hinges where these surfaces intersect. The direction of propagation of these hinge modes are determined by the sign of the $d$-wave gap, but importantly, this sign does not enter the calculation of the winding number $\nu.$
Since, for even values of $\nu$, there are an even number of hinge modes, one always tune the signs of the multiband imaginary $d$-wave order parameters such that the hinge modes form counter-propagating pairs. 
It is then possible to gap out these counter-propagating modes without changing $\nu$.  In the case of $\nu=2$,  the $d_{xy}$ order in Eq.\ \eqref{dxy} can couple the counter-propagating hinge modes and gap them without breaking the $C_4\mathcal{T}$ symmetry. By the definition of our winding number $\nu,$ the $d$-wave order does not affect it, yet it can gap the hinge states, hence we do not expect even values of $\nu$ to be stable. One can also argue that one can glue 2d chiral $p_x+ip_y$ layers to the surfaces in a $C_4\mathcal{T}$ preserving pattern which will flip the propagation directions of the hinge modes, but not destablize them. From this picture, having two copies, i.e., $\nu=2$ will not be stable since the hinge modes on one copy can be flipped and coupled to gap the original copy without breaking the symmetry.

For a weak coupling SC, the protected $\mathbb{Z}_2$ topological invariant $P=(-1)^\nu$ is given by [see Eq.\ \eqref{eq:winding2}]
\be
\label{nuparity}
P = \prod_{i}[i\sgn(\text{Re}\Delta_i)]^{C_i} = \prod_{i}[\sgn(\text{Re}\Delta_i)]^{m_i},
\ee 
where $m_i$ is the number of time-reversal invariant momentum points enclosed by the $i$-th FS, $C_i$ is the Chern number of the $i$-th FS, and $\text{Re}\Delta_i$ is understood as the $\mathcal{T}$-invariant part of the pairing gap on the FS.
In the second step we have used the properties~\cite{qi-hughes-zhang} that (i) $(-1)^{C_{i}} = (-1)^{m_i}$, and (ii) following the Nielsen-Ninomiya theorem, the total Chern number of all FS's vanishes, $\sum_{i} C_i=0$. From Eq.\ \eqref{nuparity} it is straightforward to verify that for a single-orbital spin-$\frac{1}{2}$ system, our $p+id$ state indeed is a TSC$_2$. 

The identification of the topological invariant for the 2d case is also possible. In the $\mathcal{T}$ symmetric class DIII TSC, the topological invariant is already a $\mathbb{Z}_2$ number, which indicates the presence/absence of stable  helical Majorana edge modes. The breaking of $\mathcal{T}$ with a $d$-wave order that preserves $C_4\mathcal{T}$ will generically gap the helical Majorana edge modes and generate MBS at the four corners. Thus, the topological invariant is the same as the $\mathbb{Z}_2$ number for just the $\mathcal{T}$ symmetric sector. Via a dimensional reduction procedure, it was found in Ref.~\onlinecite{qi-hughes-zhang} that in the weak pairing limit, the $\mathbb{Z}_2$ invariant can be defined as the parity of the winding number for the Hamiltonian $\mathcal{H}(k_x,k_y,\theta)$ that smoothly interpolates between the 2d SC in consideration (at $\theta=0$) and a trivial SC (at $\theta=\pi$). For a 2d $\mathcal{T}$ symmetric weak-coupling SC, the topological invariant is given by $\prod_{i}[\text{sgn}(\Delta_i)]^{m_i}$, where $i,\Delta_i, m_i$ are defined in the same way as before. Therefore, for our TSC$_2$ its topological invariant is
\begin{align}
P_{\rm 2d}=\prod_{i}\[\sgn(\textrm{Re}\Delta_i)\]^{m_i}
\end{align} where $\text{Re}\Delta_i$ is understood as the TR invariant part of the pairing gap on the FS. 

In summary we have found that the topological invariants in for both the 2d and 3d $C_4\mathcal{T}$ symmetric TSC$_2$ phases can be determined from just the $\mathcal{T}$-invariant sector. This is similar to the chiral hinge insulator with $C_4\mathcal{T}$ symmetry shown in Ref. \onlinecite{schindler2017} where the magneto-electric $\theta$-angle\cite{qi-hughes-zhang-2008} was shown to still characterize the topological phase even when $\mathcal{T}$ is broken. 
Further, by analogy, our results on the 2d TSC$_2$ topological invariant suggest that 2d $C_4\mathcal{T}$ quadrupole insulators are described by the same topological invariant as the $\mathcal{T}$-invariant quantum spin Hall insulator.


\subsection{Realization of $p+id$ pairing in a metallic system}

To realize $p+id$-wave SC order, we now explicitly construct a 2d TSC$_2$ phase with $p$- and $d$-wave pairing from an instability of a metallic normal state with electronic interactions. In the literature the pairing interactions for the $\mathcal{T}$-invariant $p$-wave SC~\cite{Fu-berg,kozii,ruhman,brydon-sau-das-sarma,wang-cho-hughes-fradkin,ruhman-savary-Lee-fu} and $d$-wave SC~\cite{Chubukov2008,palee-wen,sczhangso5} have been extensively studied.  The strategy here is to combine two types of interactions that respectively favor $p$-wave and $d$-wave order and show that by tuning the interactions to comparable strengths the system naturally develops a TSC$_2$ state with $p+id$ pairing symmetry.

Following Refs.\ \onlinecite{wang-cho-hughes-fradkin, wang-fu-2017}, a $p$-wave instability is induced by fluctuations of inversion breaking order. To this end, we consider the following interaction mediated by parity fluctuations: $H_{\rm parity} (\vec{k},\vec{k}',\vec{p},\vec{p}')= U^{\rm parity}_{\alpha\beta,\gamma\delta}(\vec{k},\vec{k}',\vec{p},\vec{p}') c^\dagger_{\alpha}({\bf k})c^\dagger_{\gamma}(\bf p) c_{\beta}({\bf k}') c_{\delta}({\bf p}')$ where
\begin{align}
\label{eq:31}
&U^{\rm parity}_{\alpha\beta,\gamma\delta}(\vec{k},\vec{k}',\vec{p},\vec{p}') \nonumber\\
&= V^{\rm parity}\left[\(\frac{\hat{\vec{k}}+\hat{\vec{k}'}}{2}\)\cdot\vec{\sigma}_{\alpha\beta}\right]\left[\(\frac{\hat{\vec{p}}+\hat{\vec{p}'}}{2}\)\cdot\vec{\sigma}_{\gamma\delta}\right],
\end{align} where $\alpha,\beta$ are (pseudo) spin indices. $V^{\rm parity}$ is the correlation function of the parity fluctuations; for our purposes we simply set it to a constant.
It is helpful to introduce the helicity operator $\chi= \hat {\bf k}\cdot \vec{\sigma}=\pm 1$, and it is straightforward to see that the scattering of electrons via this interaction preserves helicity. It is therefore convenient to introduce pairing gaps $\Delta_{\pm}(\bf k)$ on FS's with a given helicity,
\begin{align}
\label{Dpm}
H_{\pm} = \int d{\bf k} \Delta_{\pm} ({\bf k}) c^T({\bf k})  i\sigma_y P_{\pm}c(-{\bf k}) + h.c.,
\end{align}
where $P_{\pm} \equiv (1\pm\hat{\bf k}\cdot \vec{\sigma})/2$ are helicity projection operators.
For the interaction term $U_{\alpha\beta,\gamma\delta}$, the superconducting gaps $\Delta_\pm$ decouple in the linearized gap equations, though when other interactions are included $\Delta_{\pm}$ will be coupled in general. Interestingly, if $\Delta_{+}=-\Delta_{-}$ is enforced due to their coupling, then the resulting order corresponds to a $p$-wave order with $\Delta_{p}=|\Delta_{\pm}|$. Indeed we can write the p-wave pairing gap in terms of $\Delta_{\pm}$ as
\begin{align}
H_{p} =& \int d{\bf k} \Delta_{p} ({\bf k}) c^T({\bf k})  i\sigma_y \[P_+-P_{-}\]c(-{\bf k}) + h.c.\nonumber\\
= &\int d{\bf k} \Delta_{p} ({\bf k}) c^T({\bf k})  i\sigma_y (\hat{\bf k}\cdot {\vec \sigma})c(-{\bf k}) + h.c.
\end{align}

To further couple $\Delta_{\pm}$, we consider interactions that are  mediated by {antiferromagnetic fluctuations} peaked at momentum transfer ${\bf Q} = (\pi, \pi)$ with
\begin{align}
\label{eq:34}
&U^\text{af}_{\alpha\beta,\gamma\delta}(\vec{k},\vec{k}',\vec{p},\vec{p}') = V^{\rm af}\sum_{i=x,y,z}\vec{\sigma}^i_{\alpha\beta}\chi(\vec{k},\vec{k}')\vec{\sigma}^i_{\gamma\delta},
\end{align}
where the spin-spin correlation function is given by \begin{eqnarray}\begin{aligned}
\chi(\vec{k},\vec{k}')= \frac{1}{(\vec{k}-\vec{k}'-\vec{Q})^2 + \xi^{-2}}.
\end{aligned}
\label{chikk}
\end{eqnarray}
This interaction is \emph{repulsive} in nature~\cite{Chubukov2008}, and for a large enough $\xi$ favors $\Delta_{+}({\bf k})=-\Delta_{-}({\bf k+Q})$. If $U^{\rm af}$ is treated as a small perturbation, together with the dominant parity fluctuations $U^{\rm parity}$, $p$-wave order will be favored. On the other hand, if $U^{\rm af}$ is dominant over $U^{\rm parity}$, it is well-known that antiferromagnetic fluctuations by themselves favor  $d$-wave pairing. In terms of the helical pairing fields $\Delta_{\pm}$, a $d$-wave pairing order satisfies 
\begin{align}
\Delta_{d}({\bf k})=\Delta_{+}({\bf k})=\Delta_{-}({\bf k})
\end{align}
and both transforming with a sign change under a $C_4$ rotation.
Therefore, depending on the relative amplitude of $V^{\rm af}$ and $V^{\rm parity}$, either a $p$-wave order or a $d$-wave order is induced as a leading instability. We verify these claims in Appendix \ref{app:b}.

When the $p$-wave and $d$-wave instabilities are comparable, then at low temperatures the two orders can coexist. 
Again, the coexisting ground state can be determined by analyzing the Ginzburg-Landau (GL) free energy, all the symmetry-allowed terms of which are given by
\begin{align}
\label{eq:Free_energy}
\mathcal{F} =& \alpha_1 |\Delta_p|^2 +\alpha_2 |\Delta_d|^2 + \beta_1|\Delta_p|^4 + \beta_2|\Delta_d|^4\nonumber\\
& + 4\bar{\beta}|\Delta_p|^2|\Delta_d|^2 + \tilde{\beta}(\Delta_p^2\Delta_d^{*2} + \Delta_d^2\Delta_p^{*2} ),
\end{align}
where we have split the momentum dependent gaps $\Delta_{p,d}({\bf k})\equiv\Delta_{p,d}({\theta})$ into a constant part and a form-factor part, 
  i.e., $\Delta_{p,d}(\theta)=\Delta_{p,d} \times f_{p,d}(\theta)$. The form factors $f_{p,d}(\theta)$ enter the evaluation of the coefficients of the free energy. 
 
As discussed in the previous section,  $\tilde\beta$ fixes the relative phase of $\Delta_d$ and $\Delta_p$ to be $\pm\pi/2$~\cite{maiti-chubukov-2013,wang-chubukov-2014,congjun-trsb}, if they coexist. As can be verified by a straightforward minimization of the free energy, the two order parameters coexist if~\cite{wang-chubukov-2014, maiti-chubukov-2013} 
\begin{align}
\beta_1\beta_2>(2\bar\beta-\tilde\beta)^2.
\label{eq:coex}
\end{align} 
The values of the $\beta$'s can be obtained by integrating out the fermions, and are
 given by the product of fermionic Green functions and the form factors $f_{s,d}(\theta)$.~\cite{goswami-roy,wang-cho-hughes-fradkin,wang-fu-2017}
Explicitly evaluating the $\beta$'s by integrating over the fermionic Green functions, we obtain that for our circular FS,
\begin{align}
\label{eq:beta1}
&\beta_1 = \beta\int\frac{d\theta}{2\pi}f_p^4(\theta) ,
\beta_2 = \beta\int\frac{d\theta}{2\pi}f_d^4(\theta) , \nonumber\\
&\bar\beta =\tilde\beta = \beta\int\frac{d\theta}{2\pi}f_p^2(\theta)f_d^2(\theta),
\end{align}
where 
\begin{eqnarray*}\begin{aligned}
\beta = \frac{N(0)T}{2}\sum_m\int_{-\infty}^{\infty}\frac{d\epsilon}{(\omega_m^2+\epsilon^2)^2} = \frac{5\zeta(3)}{8\pi^2T^2}N(0).
\end{aligned}\end{eqnarray*}
Here $\zeta(x)$ is the Riemann zeta function. By  the Cauchy-Schwarz inequality one can prove that generally 
\begin{eqnarray}\begin{aligned}
\beta_1\beta_2>\bar{\beta}^2 = \tilde{\beta}^2,
\end{aligned}\end{eqnarray}
which we also verified numerically.  This is \emph{precisely} the coexistence condition for $\Delta_p$ and $\Delta_d$. Combined with the result on their relative phase, we have shown that the ground state has a $\mathcal{T}$-breaking $p+id$ pairing symmetry, and thus spontaneously generates a TSC$_2$ phase protected by $C_4\mathcal{T}$ symmetry.

\subsection{{Realization of $p+id$ pairing in a superconducting heterostructure}}
Alternatively, a $p+id$ pairing state can also be induced extrinsically by Josephson-coupling a $p$-wave SC and a $d$-wave SC.  In Fig.~\ref{fig:junction}(a) we illustrate such a setup of superconducting heterostructure, with, e.g., a cuprate $d$-wave SC on top, and a $p$-wave SC on the bottom.  Due to the conflicting pairing symmetries~\cite{conflicting-1,conflicting-2}, the Josephson coupling between the top and bottom layers can only be achieved by a quartic term $\sim \Delta_p^2\Delta_d^{*2} + h.c.$. Using the same argument for Eq.\ \eqref{eq:Free_energy}, the two order parameters $\Delta_p$ and $\Delta_d$ differ by a phase of $\pi/2$. By proximity effect, the cuprate layer develops $p+id$-wave order. Since the cuprate system is $C_4$ symmetric on its own, the bottom layer now realizes a $C_4{\mathcal T}$ symmetric TSC$_2$ and can host corner Majorana modes, as we illustrate in Fig.~\ref{fig:junction}(a). A similar setup was recently proposed using a heterostructure of high-$T_c$ SC and quantum spin Hall insulator~\cite{zhongwang-new,lu-zhang-2018}; there the authors found a related, proximity-induced superconducting phase with a pair of Majorana modes at each corner.

For a material realization of the $p$-wave SC layer, recent theoretical and experimental studies have identified Cu$_x$Bi$_2$Se$_3$~\cite{Fu-berg, wan-2014} and the half-Heusler compound YPtBi~\cite{paglione,ruhman-savary-Lee-fu} as promising candidates for $\mathcal{T}$-symmetric $p$-wave pairing. However, further investigations are needed to determine whether these 3d materials remain $p$-wave superconductors in a thin-film geometry.

Alternatively, we propose that one can ``mimic" a 2d $p$-wave SC using a superconducting heterostructure. Very recently it has been experimentally identified~\cite{fesete1,fesete2,fesete5} that FeTe$_{0.55}$Se$_{0.45}$ is a Fu-Kane-type~\cite{fu-kane-2008} topological superconductor with surface Dirac cones in the normal state at a rather high SC transition temperature $T_c=14.5$K. We note that, since the two Dirac cones on the opposite surfaces are of opposite helicity, the pairing gaps on them can be regarded as our $\Delta_{\pm}$ in Eq.\ \eqref{Dpm}. With a $\pi$-Josephson junction connecting the two opposite surfaces, the quasi-2d system effectively realizes a 2d $p$-wave SC. Indeed, with opposite SC gaps on the two surfaces, the SC order of the whole system is odd under spatial inversion. To generate a TSC$_2$ phase, we propose a setup based on this idea illustrated in Fig.\ \ref{fig:junction}(b). A cuprate SC thin film is sandwiched between two FeTe$_{0.55}$Se$_{0.45}$ superconductors that are connected by a $\pi$-junction. For similar reasons as above, the cuprate layer together with its interfaces with the FeTe$_{0.55}$Se$_{0.45}$ layers is in a TSC$_2$ phase that hosts four corner Majorana modes. A particularly appealing feature of this proposal is that it can potentially realize a {high-$T_c$} TSC$_2$.}

\begin{figure}
\centering
\includegraphics[width=\columnwidth]{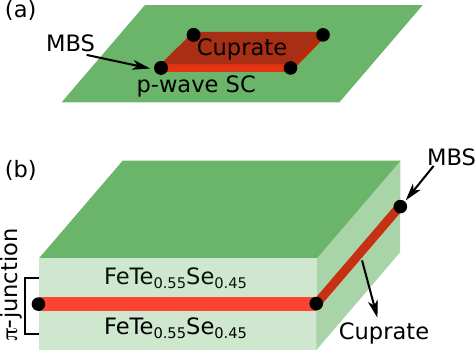}
\caption{Two experimental setups for a proximity-induced $\text{TSC}_2$ phase. (a) A cuprate superconductor with $d$-wave symmetry is placed on top of a $\mathcal{T}$-invariant $p$-wave superconductor. The Josephson coupling induces a quartic term in the free energy, forcing the two order parameters to differ by a $\pm\pi/2$ phase. Due to the $C_4$ symmetry of the cuprate system, it develops a $C_4\mathcal{T}$-invariant $p+id$ wave order with MBS's at the corners. (b) A cuprate superconductor is sandwiched between two iron-based superconductors $\text{FeTe}_{0.55}\text{Se}_{0.45}$. The cuprate layer, together with its interfaces with $\text{FeTe}_{0.55}\text{Se}_{0.45}$ layers, realizes a 2d TSC$_2$, with four MBS's at the corners.}
\label{fig:junction}
\end{figure}

\section{Conclusion}
In this work, we have studied 2d and 3d 2nd order topological superconductors, which host Majorana bound states at the corners in the 2d system, and gapless, chiral Majorana modes at the hinges of a 3d system. The purpose of this work was twofold: to understand the topological properties TSC$_2$ such as their symmetry requirements and topological invariants, and to investigate how these exotic superconducting states may be realized in weak-pairing scenarios.

We have identified two routes towards TSC$_2$ phases The first route is through inducing  $p_x+ip_y$ order on a 2d Dirac semimetal with four mirror-symmetric Dirac nodes. Such a band structure can be realized either in a magnetic, or spin-polarized two-band electronic system or in cold atom systems~\cite{demler2017}. The intrinsic particle hole symmetry quantizes the $\mathbb{Z}_2$ topological invariant defined in Eq.~\eqref{eq:def_qxy}; With mirror symmetry, the invariant can be expressed through nested Wilson loops. Furthermore, we have shown that in the presence of a chemical potential $\mu$, a finite-range, attractive interaction naturally induces such a $p_x+ip_y$-wave pairing in the doped Dirac point normal state. 

For our second system we considered a somewhat more exotic $C_4\mathcal{T}$-symmetric $p+id$ order, but in this scenario the requirement on the normal state is much less restrictive, i.e., just a featureless, spin-degenerate  Fermi surface. Remarkably, we have shown that the topological invariant of this class of TSC$_2$ is $\mathbb{Z}_2$ in both 2d and 3d, and were able to express the topological invariants in simple formuale involving just the low-energy properties of the system.
We have found that a combination of interactions favoring $p$-wave and $d$-wave orders naturally induces the $p+id$ pairing symmetry to generate the TSC$_2$ phase. Alternatively, we proposed that the $p+id$ pairing order may also be induced by proximity effect in a superconducting heterostructure, which can even potentially realize a \emph{high-$T_c$} TSC$_2$ system.

One interesting extension of the present work is whether TSC$_3$'s, which are 3d topological superconductors with eight vertex modes, can be realized.  Building from a Majorana plaquette model similar to an octupole version of Eq.\ \eqref{H0}, it is not difficult to construct a BdG Hamiltonian for TSC$_3$ for a four-band normal state. However, we did not find an analogous identification like the TSC$_2$ case where the BdG Hamiltionian describes a superconducting order that develops from a gapless band structure. Thus we do not expect that TSC$_3$'s can be spontaneously realized by simply generalizing the analysis in Sec.\ \ref{sec:mirror}. We leave the issue of realizing TSC$_3$ to future work.

\begin{acknowledgments}
We thank W. A. Benalcazar and Zhong Wang for discussions.
YW acknowldeges support from the Gordon and Betty Moore Foundations EPiQS Initiative through Grant No. GBMF4305. ML thanks NSF Emerging Frontiers in Research and Innovation NewLAW program Grant EFMA-1641084 and NSF CAREER Grant DMR-1351895 for support. TLH was supported by the ONR YIP Award N00014-15-1-2383.
\end{acknowledgments}

\appendix

\begin{figure}[htbp]
\centering
\includegraphics[width=\columnwidth]{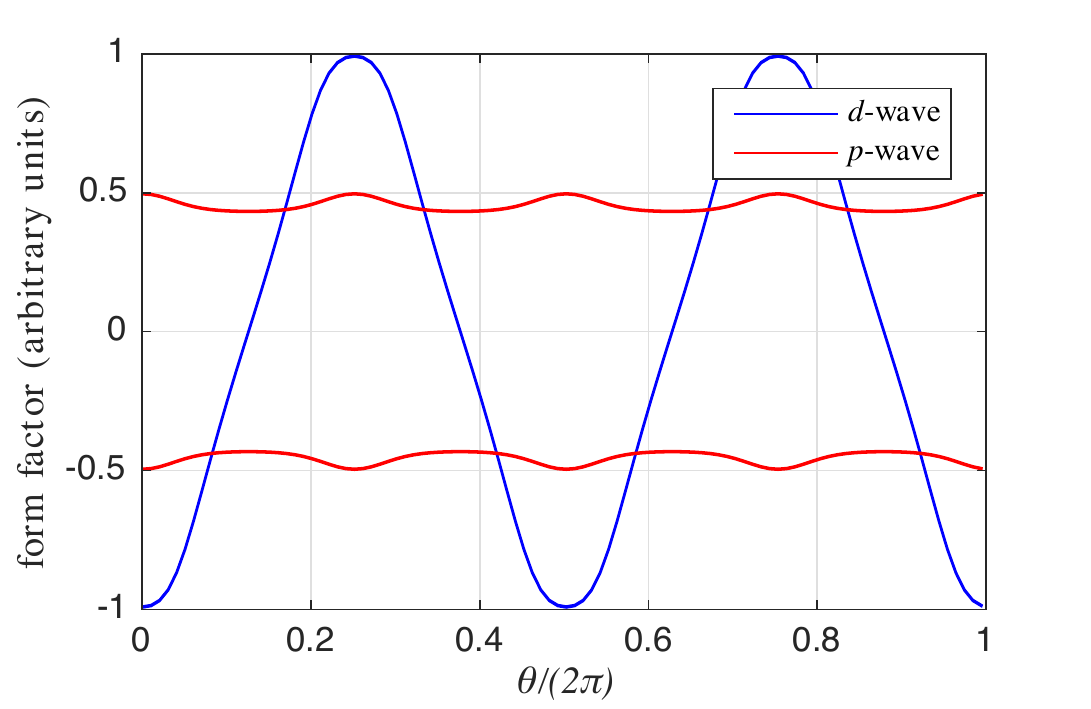}
\caption{The form-factors of the superconducting pairing functions for the two leading instabilities. We find the two leading instabilities are $p$-wave (red) and $d$-wave order. For $p$-wave order the two red curves are for the gap functions $\Delta_{\pm}(\theta)$ separately, and we can see the relative sign change that is characteristic of $p$-wave order. 
 The d-wave pairing has the characteristic four node structure at $\theta=\pm\pi/4,\pm3\pi/4$. To ensure that the Fermi surface is large enough to allow for a $\bf Q$ momentum transfer, we chose $k_F=5\pi/(6a_0)$. We have set the antiferromagnetic correlation length $\xi=1,$ and we have verified that choosing other values does not change our results.}
\label{fig:for_p_d}
\end{figure}

\section{The absence of a Wannier transition in $\mathcal{H}'({\bf k},\alpha)$}
\label{app:a}

In this Appendix we show that $\mathcal{H}'({\bf k},\alpha)$, defined as a smooth interpolation between Eq.\ \eqref{eq:1} (at $\alpha=0$)and Eq.\ \eqref{eq:bdg'} (at $\alpha=1$) does not go through a Wannier transition, i.e., gap closing for the edge Hamiltonian as a function of $\alpha$.
{From mirror symmetry this could only occur through a band inversion of the edge Hamiltonian at high symmetry points, say $k_{x,y}=0$ or $\pi$. 
To confirm this does not happen, we need to show that the 1d subsystem $h(k_y,\alpha)\equiv \mathcal{H}'(k_x=0, k_y,\alpha)$ at $k_x=0$, when treated as an effective 1d superconductor with boundaries in the $y$-direction can never host boundary zero modes throughout this deforming process. This 1d Hamiltonian $h(k_y,\alpha)$ can again be split into a normal-state part and a pairing-gap part.
Since the Dirac points are always in the four quadrants, the normal state spectrum of $h(k_y,\alpha)$ with eigenvalues $e_{\pm}(k_y,\alpha)=\pm \sqrt{f_1^2(0,k_y,\alpha)+f_2^2(0,k_y,\alpha)}$ is gapped. The two bands have opposite spin texture, and  importantly, our same-spin, $p$-wave pairing for this subsystem only pairs within each of the two bands. Then the subsystem, {throughout this process}, is just two copies of \emph{decoupled} $p$-wave SC.  In the Nambu space of each given band, the effective Hamiltonian is
\begin{align}
h_{\pm}(k_y,\alpha)=& \pm \sqrt{f_1^2(0,k_y,\alpha)+f_2^2(0,k_y,\alpha)} s_z + \nonumber\\
+&\sgn(k_y)\sqrt{g_1^2(0,k_y,\alpha)+g_2^2(0,k_y,\alpha)} s_x
\end{align}
where $s_{z,x}$ are Pauli matrices in the Nambu subspace. (Note that $\sgn(k_y)$ is enforced by Fermi statistics, and for odd $g_{1,2}$, the pairing gap is a smooth function of $k_y$.) From the well-known results in early works~\cite{readgreen,kitaev2001}, it is clear that both $h_{\pm}$ are in the trivial phase, {because the normal state spectrum does not host a Fermi surface (``strong pairing phase" in the terminology of Ref.\ \onlinecite{readgreen}).}
Thus $h_{\pm}(k_y)$ do not host boundary zero modes. Since $h(k_y,\alpha)$ decouples into $h_{\pm}(k_y)$, it does not host boundary zero modes either. 
The same arguments can be applied to other high-symmetry subsystems at $k_x=\pi$ and $k_y=0,\pi$. This way we have proven that Eq.\ \eqref{eq:bdg'} with $\mu=0$ is topologically equivalent with Eq.\ \eqref{eq:1}. We can now turn on a nonzero chemical potential $\mu$ term. For sufficiently small $\mu$ that is smaller than the bandwidth, the topology of $h_{\pm}$ do not change, and there is no gap closing either at the edge or in the bulk. 
With no gap closing, the particle-hole symmetry is sufficient to protect the corner MBS's.}

\section{Linear gap equation for $p$-wave and $d$-wave pairing}
\label{app:b}
{In this Appendix we verify that the combined interactions given by Eqs.~(\ref{eq:31}) and (\ref{eq:34}) lead to instabilities towards $p$-wave and $d$-wave order parameters.} We consider a circular FS parametrized by an angle $\theta$, where 
\begin{align}
\vec{k} &= (k_F\cos\theta,k_F\sin\theta),
\end{align}
where $k_F$ is the Fermi momentum. 
After summing over the spin indices for parity and antiferromagnetic fluctuations, and within the BCS approximation, the linear gap equations for $\Delta_{\pm}$ parametrized by the FS angle $\theta$ are given by
\begin{widetext}
\begin{align}
\label{eq:spin_singlet}
{\lambda_c}\Delta_+(\theta) =& \int d\theta' \[V^{\rm parity}\cos^2\left(\frac{\theta-\theta'}{2}\right)\Delta_{+}(\theta')-V^\text{af}\chi(\theta,\theta')\(\frac{3-\cos(\theta-\theta')}{2}\Delta_+(\theta')+\frac{3+\cos(\theta-\theta')}{2}\Delta_-(\theta')\)\]\nonumber\\
\lambda_c\Delta_-(\theta) =& \int d\theta' \[V^{\rm parity}\cos^2\left(\frac{\theta-\theta'}{2}\right)\Delta_{-}(\theta')-V^\text{af}\chi(\theta,\theta')\(\frac{3-\cos(\theta-\theta')}{2}\Delta_-(\theta')+\frac{3+\cos(\theta-\theta')}{2}\Delta_+(\theta')\)\],
\end{align}
\end{widetext}
where $\lambda_c=1/[N(0)\log\frac{\Lambda}{T_c}]$, $N(0)$ is the density of states at the Fermi level, and $\chi(\theta,\theta')$ is $\chi({\bf k},{\bf k}')$ in Eq.\ \eqref{chikk} projected to the Fermi surface. {The factors $\cos^2[(\theta-\theta')/2]$ and $[3-\cos(\theta\pm\theta')]/2$ comes from the product of projection operators $P_{\pm}({\bf k})$ and the spin dependence of $U^{\rm parity}$ and $U^{\rm af}$. }
By Fermi statistics, we explicitly only keep solutions that satisfy $\Delta_{\pm} (\theta) =\Delta_{\pm}(\theta+\pi)$. (Note that even the odd-parity $p$-wave order satisfies this.) 
This set of linear integral equations can be solved numerically as an eigenvalue problem in the vector space of $[\Delta_+(\theta), \Delta_-(\theta)]$. From the eigenvalue $\lambda_c$ one can obtain the mean-field critical temperature $T_c$ of the pairing channels. The eigenfunctions for $\Delta_{\pm} ({\bf k})$ with the largest $\lambda_c,$ and thus highest $T_c$'s, correspond to channels of strongest pairing instability.
Indeed, as we expected from the heuristic arguments above, with this combination of interactions, the two leading pairing instabilities are towards $d$-wave and $p$-wave (as is confirmed in Fig.\ \ref{fig:for_p_d} in which we plot the two eigenfunctions (pairing form-factors) that had the largest eigenvalues). 
By tuning $V^{\rm parity}$ and $V^{\rm af}$, either $p$-wave or $d$-wave is dominant. For our model, when $V^{\rm af}={1.234}V^{\rm parity}$, the two instabilities are degenerate.

\bibliographystyle{apsrev4-1}
\bibliography{2ndTSC}

\end{document}